\documentclass[aip,reprint,amsmath,amssymb]{revtex4-2}
\usepackage{booktabs,multirow,amsmath}
\usepackage{color}
\usepackage{graphicx}
\usepackage{amsmath}
\usepackage{subcaption}
\usepackage{amssymb}
\usepackage{lineno}
\usepackage{natbib}

\graphicspath{{figs/}}

\usepackage{placeins}

\begin{document}
	\title{Plasma transport by carrier electrons: classical coefficients across ionization and magnetization}
	
	\author{C. A. Walsh}
	\email{walsh34@llnl.gov}
	\affiliation{Lawrence Livermore National Laboratory, 7000 East Avenue, Livermore, CA 94550}

	\date{\today}
	

	\begin{abstract}
		This paper explains the behavior of the plasma transport coefficients by studying the electron population carrying each flux. Both electron-electron (e-e) collisions at low $Z$ and magnetization affect fast electrons preferentially; transport carried by faster electrons is therefore more suppressed by these effects than transport carried by slower electrons. An analytic single-carrier model motivates measurement of a carrier Hall parameter $\psi=\chi f(Z)$ for each process; the presented fits allow magnetization to be characterized independent of ionization. A two-carrier model represents classical transport more faithfully than a single carrier, by separating the population into two opposing streams. The coefficients then reduce to differences between the streams: in energy, rotation and drive (how they respond to $\nabla T_e$ relative to $\mathbf{E}$). Nernst is reduced at low-$Z$ more than thermal conductivity due to e-e collisions reducing relative B-field rotation of the two populations. Adding a third population (a slow drifting Maxwellian) completes the model by providing a means for electrons to pass their momentum to the ions. This simple model predicts every coefficient's dependence on ionization and magnetization to within about $10$--$20\%$, with nothing fitted. This is the first paper in a 4-part series; the subsequent papers extend the carrier picture to kinetic heat-flux (paper 2), kinetic magnetic field advection (paper 3), and kinetic magnetic field generation (paper 4).
	\end{abstract}
	\maketitle
	
	Magnetic fields are ubiquitous in high-energy-density plasmas \cite{lezhnin2026,haines1981,mason1998,sarri2012,fiksel2014,walshSelfGeneratedMagneticFields2017,schoeffler2014}. These fields modify thermal conduction down temperature gradients\cite{braginskiiTransportProcessesPlasma1965}as well as initiating transport of electron energy with the electron current. Likewise, the temperature gradients and currents transport magnetic field\cite{hainesHeatFluxEffects1986}. The magnitude of these processes is governed by the coveted transport coefficients. 
	
	Each transport process (and corresponding coefficient) is dependent on a particular range of electron energies. Heat-flux, for example, is carried by higher energy electrons than the electrons resistively diffusing the magnetic field. This paper studies how the properties of the transport coefficients (their dependence on ionization $Z$ and magnetization $\chi$) emerge from the velocity of the carrier electrons. 
	
	The classical transport of magnetic fields in a plasma can be written as follows: \cite{epperleinPlasmaTransportCoefficients1986,walshExtendedmagnetohydrodynamicsDensePlasmas2020,sadlerMagneticFieldGeneration2020, sadlerSymmetricSetTransport2021}
	\begin{align}
		\begin{split}
			\frac{\partial \mathbf{B}}{\partial t} = &\nabla \times \Bigg( \frac{\nabla P_e}{e n_e} - \frac{\beta_{\parallel} \nabla T_e}{e}\Bigg) \\ 
			& - \nabla \times \frac{\eta}{\mu_0 } \nabla \times \mathbf{B} \\
			&+ \nabla \times \Big((\mathbf{v}_{plasma}+\mathbf{v}_{Nernst}+\mathbf{v}_{j,B}) \times \mathbf{B} \Big)
			\label{eq:mag_trans_new}
		\end{split}
	\end{align}
	The first line are sources of magnetic flux; Biermann battery generation and composition Sadler generation at ionization gradients \cite{sadlerMagneticFieldGeneration2020}. The second line is  diffusion of magnetic field with diffusivity $\eta=\alpha_{\parallel}/(\mu_0 e^2 n_e^2)$ .
	
	The final line in equation \ref{eq:mag_trans_new} is advection of magnetic field with the bulk plasma, temperature gradient (Nernst) and current:
	
	\begin{equation}
		\label{eq:mag_trans_vnernst}	\mathbf{v}_{Nernst} = - \gamma_{\bot} \nabla T_e - \gamma_{\wedge}\hat{\mathbf{b}} \times \nabla T_e 
	\end{equation}
	
	\begin{equation}
		\label{eq:mag_trans_vjB}	\mathbf{v}_{j,B} =  - \frac{\mathbf{j}}{e n_e}(1 + \delta_{\bot}^c) + \frac{\delta_{\wedge}^c}{e n_e} (\mathbf{j} \times \hat{\mathbf{b}})   
	\end{equation}
	
	The electron energy is also transported by analogous processes: down temperature gradients and with the electron current. The magnetized thermal conduction equation is:
	\begin{equation}
		\mathbf{q}_e = -\kappa_{\parallel} \nabla_{\parallel}T_e -\kappa_{\bot} \nabla_{\bot}T_e - \kappa_{\wedge} \hat{\mathbf{b}} \times \nabla T_e  \label{eq:magheatflow}
	\end{equation}
	While the advection of electron energy with current is:
	\begin{equation}	
		\mathbf{v}_{j,T_e} =  - \frac{\mathbf{j}_\parallel}{e n_e}(1 + \beta_{\parallel}^c)- \frac{\mathbf{j}_\bot}{e n_e} (1 + \beta_{\bot}^c) +\frac{\beta_{\wedge}^c}{e n_e} (\mathbf{j} \times \hat{\mathbf{b}})  \label{eq:magtrans_heat_j}
	\end{equation}

	\begin{figure}[tb]
		\includegraphics[width=\linewidth]{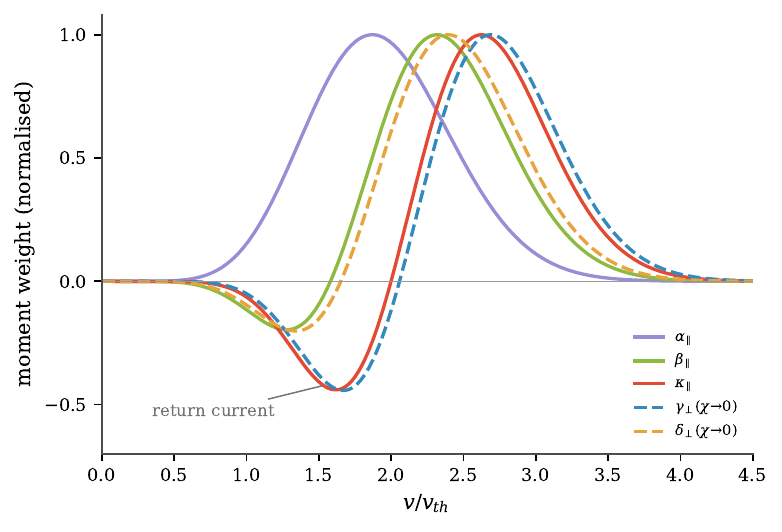}
		\caption{The velocities responsible for each transport coefficient for an unmagnetized ($|\mathbf{B}|=0$) Lorentzian ($Z=\infty$) plasma. The heat-flux ($\kappa_{\parallel}^c$) is transported by a higher velocity electron population than is responsible for resistive diffusion ($\alpha_{\parallel}$). \label{fig:lorentz_weights} }
	\end{figure}
	
	Equations \ref{eq:mag_trans_new}, \ref{eq:mag_trans_vnernst}, \ref{eq:mag_trans_vjB}, \ref{eq:magheatflow} and \ref{eq:magtrans_heat_j} are a physically intuitive way of writing classical plasma transport; every term has a defined role in advecting magnetic field (down temperature gradients or with electron current), diffusing magnetic field by resistivity, sourcing magnetic field, diffusing electron energy or advecting electron energy with the electron current. The full set of \emph{dimensionless} coefficients here are $\underline{\underline{\alpha}}^c$, $\underline{\underline{\gamma}}^c$, $\underline{\underline{\delta}}^c$,  $\underline{\underline{\kappa}}^c$ and $\underline{\underline{\beta}}^c$. 	How these coefficients depend on plasma ionization and magnetization are the subject of this paper. 
	
	The dimensional forms are then constructed using:
	$\underline{\underline{\alpha}} = \underline{\underline{\alpha}}^c m_e n_e / \tau_{B}$, $\underline{\underline{\beta}} = \underline{\underline{\beta}}^c$, $\underline{\underline{\kappa}} = \underline{\underline{\kappa}}^c n_e T_e \tau_{B} / m_e$, $\underline{\underline{\gamma}} = \underline{\underline{\gamma}}^c \tau_{B}/m_e$, $\underline{\underline{\delta}} = \underline{\underline{\delta}}^c/(e n_e)$ \cite{epperleinPlasmaTransportCoefficients1986}. $\tau_B$ is Braginskii's definition of the electron-ion (e-i) collision time\cite{braginskiiTransportProcessesPlasma1965} $\tau_B=3\sqrt{\pi}\tau_{ei}/4$, with $\tau_{ei}=\lambda_{ei}/v_{th}$. The velocity of a thermal electron is $v_{th}=\sqrt{2T_e/m_e}$. The e-i mean free path definition used is:
	\begin{equation}
		\lambda_{ei} \;=\;  \frac{16 \pi \epsilon_0^2 T_e^2}{n_e Z e^4 \ln\Lambda}
		\label{eq:lambdaei}
	\end{equation}

	The $\underline{\underline{\gamma}}^c$ and $\underline{\underline{\delta}}^c$ coefficients are directly related to $\underline{\underline{\beta}}$ and $\underline{\underline{\alpha}}$: $\gamma_{\bot}^c = \beta_\wedge^c/\chi$, $\gamma_{\wedge}^c = (\beta_{\parallel}^c-\beta_\bot^c)/\chi$, $\delta_\bot^c = \alpha_{\wedge}^c/\chi$, $\delta_{\wedge}^c = (\alpha_{\parallel}^c-\alpha_\bot^c)/\chi$. \cite{walshExtendedmagnetohydrodynamicsDensePlasmas2020}
	
	The symmetry in transport of magnetic field and electron heat in a plasma is already clear by comparing equation \ref{eq:mag_trans_vnernst} with \ref{eq:magheatflow} and equation \ref{eq:mag_trans_vjB} with \ref{eq:magtrans_heat_j}.

	The coefficients have parallel ($\parallel$), perpendicular ($\bot$) and cross ($\wedge$) components. The first indicates transport along magnetic field lines; this is unaffected by magnetization. Note that transport of magnetic field along field lines does not change the magnetic field strength, and so no parallel magnetic transport coefficients appear in the above equations.
	
	For no magnetic field, the perpendicular coefficients equal their parallel counterparts. As magnetization increases, the electrons begin undertaking Larmor orbits with radius $r_L = m_e v_\bot / e|\mathbf{B}|$. Small orbits lower transport across field lines, thereby reducing the perpendicular coefficients. The Hall parameter $\chi = \omega_e \tau_B$ is the typical metric for magnetization, with $\omega_e = e|\mathbf{B}|/m_e$ the electron gyrofrequency. $\chi = 1$ corresponds to an electron at velocity $1.10v_{th}$ turning 1 radian between collisions, due to the use of $\tau_B=3\sqrt{\pi}\tau_{ei}/4$ in the definition of $\chi$. As will be seen later, using this velocity of  electron to characterize plasma magnetization can greatly under-represent the magnetization of transport processes that are actually carried by higher velocity electrons.
	
	The perpendicular fluxes lower with magnetization due to two processes: the distance between collisions tending towards $r_L$ and the flux being re-directed by the field. The cross coefficients are this second part: the  rotated flux that is perpendicular to both the driver ($\nabla T_e$ or $\mathbf{j}$) and the magnetic field unit vector $\hat{\mathbf{b}}$. The cross terms are zero at $\chi=0$ due to no rotation mechanism. At large magnetization they decrease again due to $r_L$ decreasing. The cross terms peak at some moderate magnetization. Imagine heat-carrying electrons rotating 90 degrees between orbits, transporting heat perpendicular to $\nabla T_e$; this is the Righi-Leduc heat-flux $\kappa_{\wedge}^c$. 
	
	
	All plasma transport comes from these simple pictures of individual electrons carrying heat and magnetic field. But which electrons do the work? Generally speaking it is the higher energy electrons, although there are fewer of them. Figure \ref{fig:lorentz_weights} shows the relative importance of each velocity for each coefficient in a Lorentz ($Z=\infty$) unmagnetized ($|\mathbf{B}|=0$) plasma. The thermal conductivity $\kappa$ depends on considerably higher velocity electrons than the resistivity $\alpha$. The curves show which are the \emph{carrier electrons}. How the carriers of each transport process change with finite $Z$ (due to collisions) and B-field (due to magnetization) will be discussed throughout this work.
	
	
	
	This paper is \emph{not} intended to improve the coefficient fits once again\cite{braginskiiTransportProcessesPlasma1965,epperleinPlasmaTransportCoefficients1986,daviesTransportCoefficientsMagneticfield2021,sadlerSymmetricSetTransport2021} (although the author points to the last two citations as the most up-to date, which are essentially equivalent \cite{walsh2021magnetized}). This paper's purpose is to provide physical insight into the nature of the coefficients as ionization and magnetization are varied. Why do electron-electron collisions (which dominate at $Z=1$) modify each coefficient differently? Why is the impact of magnetization different for each ionization? Braginskii obtained the coefficients by expanding the electron distribution in a few Laguerre polynomials\cite{braginskiiTransportProcessesPlasma1965}; section \ref{sec:shapes} returns to this approach, with the expansion functions chosen for their physical meaning.
	
	The competing magnitudes of the transport processes are of particular interest. Take the example of heat-flux (energy moving down temperature gradients) and Nernst (magnetic field moving down temperature gradients). These are linked processes: both depend on $\nabla T_e$ \emph{and} modify $\nabla T_e$, with Nernst acting to demagnetize the hot plasma. Transport at a thermal front is therefore highly dependent on the ratio of these coefficients.
	
	Section \ref{sec:mag_and_ee} provides the physical underpinning of the paper, showing how magnetization and electron-electron collisions modify the electron distributions. Section \ref{sec:ideal} provides a simplistic model for transport based off a single population of electrons carrying the flux; this suggests  processes carried by higher speed electrons magnetize easier, which is borne out through fits for the carrier Hall parameters. However, the single-carrier approach violates a key principle of transport: the net particle flux does not equal the prescribed current. 
	
	Section \ref{sec:two_carrier} improves the model by introducing two carrier populations with opposing fluxes. The coefficients reduce to gaps in energy, drive and rotation between the two carriers. This explains the relative $Z$-dependence of the coefficients. Section \ref{sec:shapes} closes by introducing a drifting Maxwellian to the model, which is a sink of momentum for the electron population. With it, every coefficient's dependence on ionization and magnetization is predicted to within $10$--$20\%$ with nothing fitted.
	
	Appendix \ref{app:moments} writes the electron distribution moments used to obtain the coefficients. Appendix \ref{app:carrier} presents fits to the carrier-averaged velocity for each process against $Z$ and $\chi$. Appendix \ref{sec:ee_colls} demonstrates the importance of the full electron-electron collision operator for calculating transport.

	\section{Effect of magnetization and electron-electron collisions on the electron velocity distribution \label{sec:mag_and_ee}}
	
	\begin{figure}[tb]
		\includegraphics[width=\linewidth]{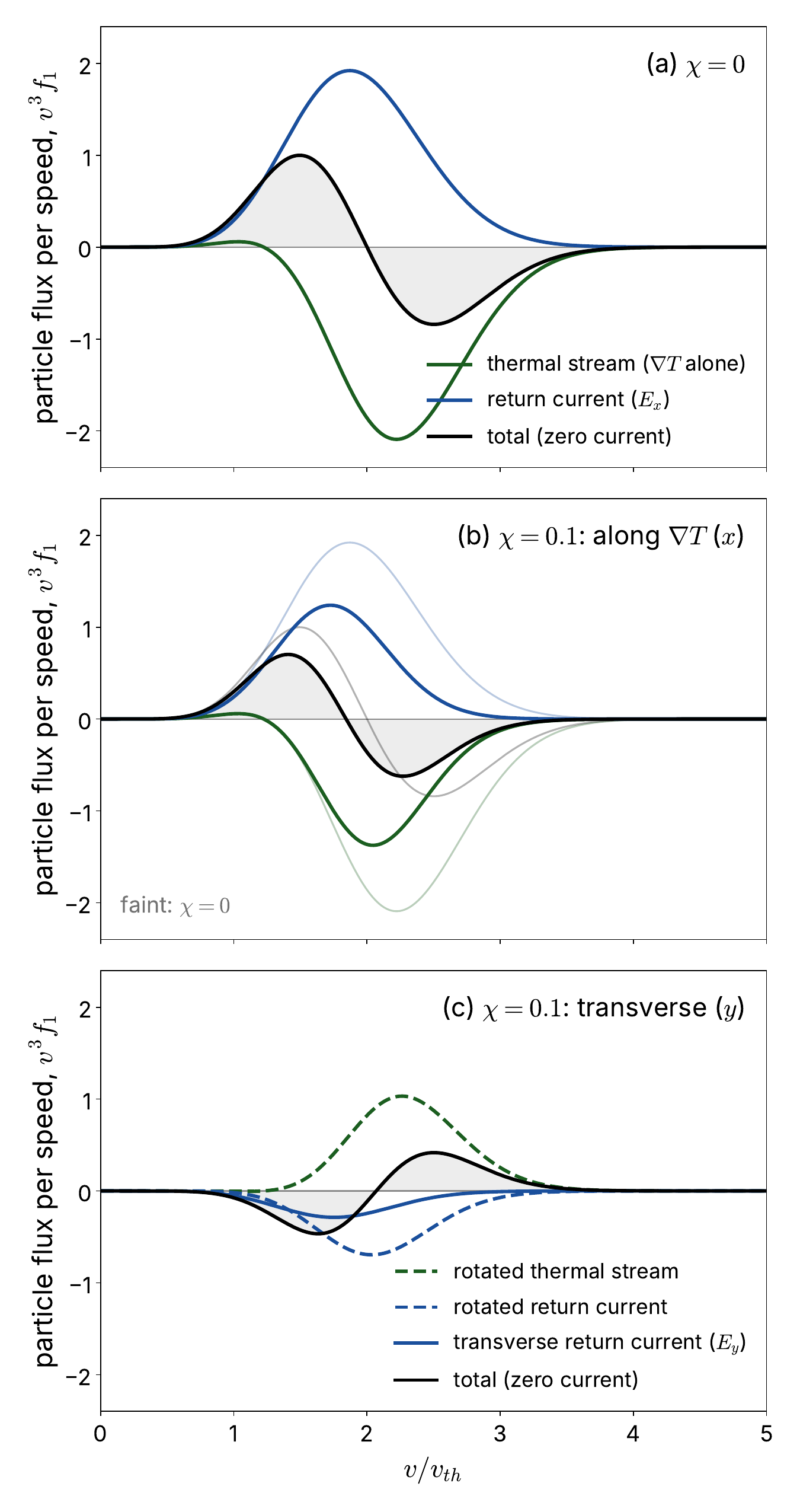}
		\caption{\label{fig:streams} (a) Particle flux $v^3f_{1x}$ for an unmagnetized Lorentz ($Z=\infty$) plasma with a temperature gradient along $x$ and no bulk current. The thermal stream driven by $\nabla T_e$ (green), the return current driven by the electric field $E_x$ (blue) and their sum (black), which has zero area. (b) The same at $\chi=0.1$. (c) The $y$-component at $\chi=0.1$. The rotated thermal stream and rotated return current (dashed) do not cancel, so a transverse return current driven by $E_y$ (solid blue) is needed to restore zero current. $E_y$ is the field that drives Nernst advection.}
	\end{figure}
	
	For a classical plasma, the electron velocity distribution can be decomposed into two pieces: the isotropic $f_0(v)$ and the anisotropic $\mathbf{f}_1(v)$.\cite{bell1981} $f_0$ is simply how many electrons are travelling at speed $v$; the moments of $f_0$ give bulk plasma properties such as electron density and temperature. $\mathbf{f}_1$ is where transport properties live. 	$f_{1x}(v=v_1) > 0$ means more electrons with $v=v_1$ are moving in positive $x$ than negative $x$. 
	
	There are two steps to finding the classical coefficients. The first is to calculate $\mathbf{f}_1$, the origin of all coefficients. The second is to take the appropriate moment of $\mathbf{f}_1$ for each coefficient. It is only after taking the moments of $\mathbf{f}_1$ that the impact of magnetization and e-e collisions can be assessed.
	
	Two terms drive transport fluxes: a bulk electron current and a temperature gradient. Both must be known to obtain $\mathbf{f}_1$. 
	
	Figure \ref{fig:streams}(a) plots the particle flux $v^3f_{1x}$ for an unmagnetized $Z=\infty$ plasma with a current along $x$ and a temperature gradient along $x$. If there is no temperature gradient, then $f_{1x}$ is found by satisfying the blue curve alone. If there is no bulk current, the temperature gradient drives a thermal stream (green line in figure \ref{fig:streams}(a)). An electric field $E_x$ (blue line) forms to satisfy the zero current condition; the total particle flux (black line) gives $f_{1x}$, which has zero area.  
	
\begin{figure}[tb]
		\includegraphics[width=\linewidth]{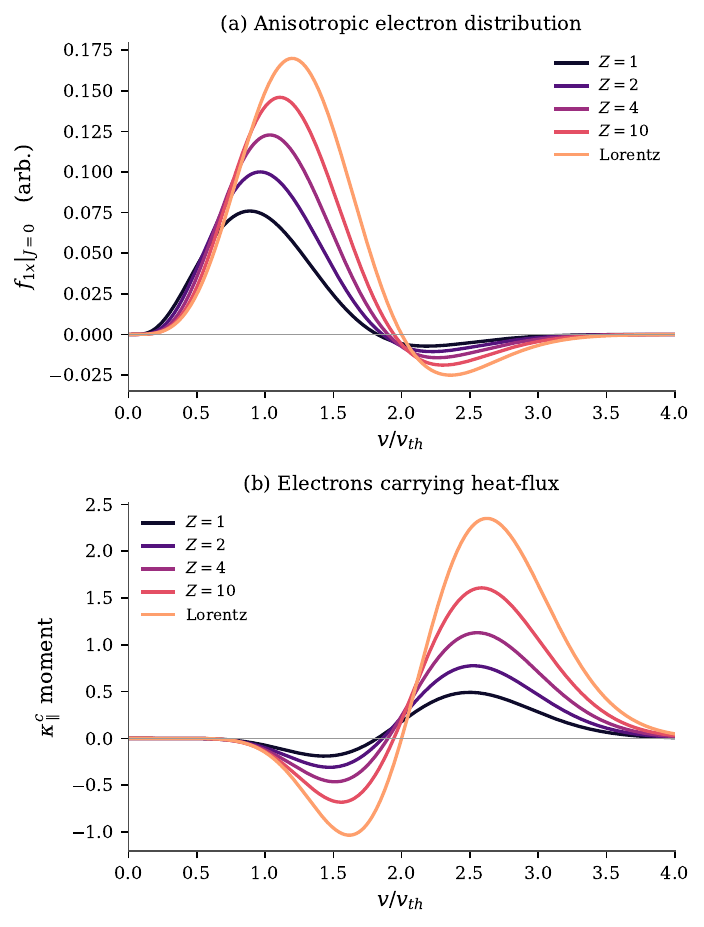}
		\caption{\label{fig:ee_on_f1} (a): Anisotropic distribution of electron velocities ($f_{1x}$) for an unmagnetized $Z=1$ plasma with $\partial T_e / \partial x > 0$ and no net current. (b): Heat-flux moment for the same plasma. e-e collisions act to reduce the thermal conductivity (area under the curve in (b)) as well as reduce the heat-flux carrier velocity (shift to the left in (b)).}
	\end{figure}

	Figure \ref{fig:ee_on_f1} continues the process of calculating an unmagnetized coefficient, now specific to transport driven by $\partial_x T_e$. Figure \ref{fig:ee_on_f1}(a) is $f_{1x}$ for a range of $Z$. There is no bulk current in this plasma ($\int f_{1x} v^3 dv = 0$). The temperature gradient drives anisotropy in the high energy electron population, while the low energy electrons (on average) travel in the opposite direction to conserve zero bulk current. For a Lorentzian $Z=\infty$ plasma the collisions are dominated by the ions. As $Z$ decreases, the electron-electron (e-e) collisions also play a role, helping to remove the velocity anisotropy driven by the temperature gradient.
	
	$f_{1x}$ from figure \ref{fig:ee_on_f1}(a) is a general result for a plasma in a temperature gradient; it is not specific to heat-flow. Figure \ref{fig:ee_on_f1}(b) demonstrates how thermal conductivity is calculated, by taking the heat-flux moment of $f_{1x}$. The heat-flux moment ($\kappa_{\parallel}^c$) is found by taking the area under the curve. e-e collisions act to reduce the heat-flux moment, thereby reducing the thermal conductivity. The collisions preferentially isotropize the higher energy population, lowering the average carrier electron speed. 
	
	To find the magnetized thermal conductivity, $f_{1x}$ must be found again. Figure \ref{fig:streams}(b) shows how the particle flux from figure \ref{fig:streams}(a) looks once the plasma is magnetized to $\chi=0.1$. Again, applying a temperature gradient and enforcing zero bulk current gives both an electric field $E_x$ and a total particle flux. Magnetization can be quantified for \emph{individual} electron velocities as:	
	\begin{equation}
		\chi(v) = \frac{\omega_e}{\nu_{ei}(v)}
		= \frac{4}{3\sqrt{\pi}}\,\left(\frac{v}{v_{th}}\right)^3 \chi,
		\label{eq:chi_v}
	\end{equation}
	Note that higher velocity electrons are more magnetized; this is apparent in figure \ref{fig:streams}(b), with the higher energy portion of $f_{1x}$ depleted more by the magnetic field than the lower energies. The field acts to remove the preferred direction ($-x$) of the electrons, thereby lowering the particle flux. 
	
	While some of that flux is truly isotropized, at moderate magnetization it is also redirected into the $y$-direction. Figure \ref{fig:streams}(c) shows $f_{1y}$; the dashed lines are the rotated thermal stream and the rotated return current that originated in $x$. Zero current is still enforced; the rotated streams, which depend on different electron velocities and are therefore rotated by different amounts, no longer cancel. Therefore, an $E_y$ field forms to set zero current once again; this is the electric field that drives Nernst advection of magnetic fields. 
	
\begin{figure}[tb]
		\includegraphics[width=\linewidth]{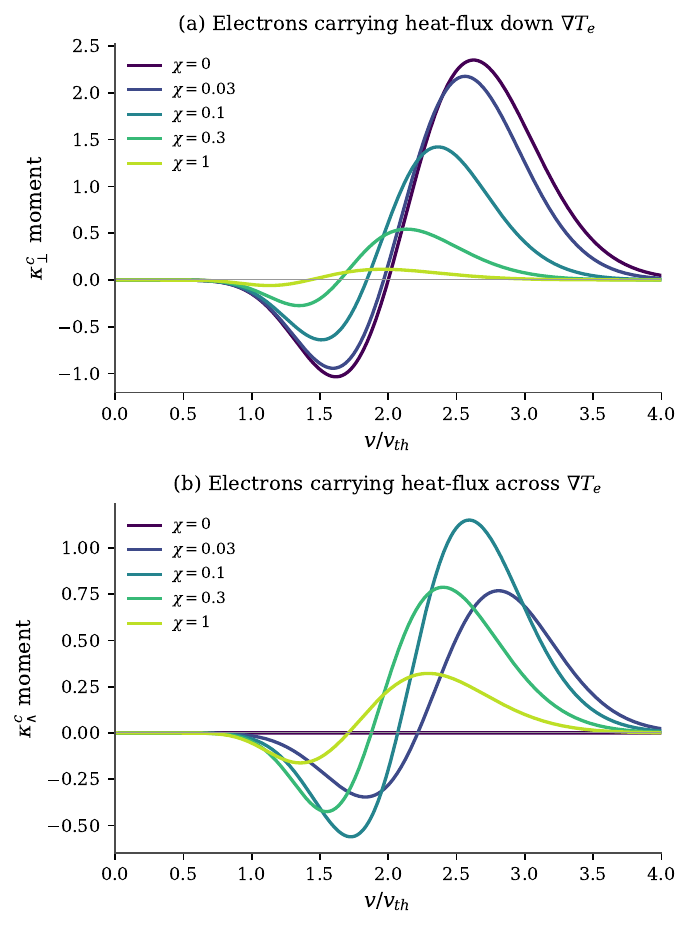}
		\caption{\label{fig:field_on_f1} Lorentzian heat-flux moments for $\kappa_{\bot}^c$ (a) and $\kappa_{\wedge}^c$ (b) for varying magnetization. Faster electrons are preferentially magnetized, resulting in lower carrier velocities with higher magnetic field.  }
	\end{figure}

	Figure \ref{fig:field_on_f1}(a) shows the Lorentzian heat-flux moment ($\kappa_{\bot}^c$) for various magnetizations. The magnitude is decreased as the magnetic field increases due to the electrons traveling less far between collisions. The higher velocity electrons are preferentially magnetized due to the $(v/v_{th})^3$ dependence in equation \ref{eq:chi_v}, resulting in the peak shifting to lower velocity.
	
	Figure \ref{fig:field_on_f1}(b)	shows the $\kappa_{\wedge}^c$ weight on $f_{1y}$ for the same plasma as \ref{fig:field_on_f1}(a) to demonstrate the same concept for the cross coefficients. In this case finite magnetization is required to rotate the anisotropy in $f_{1x}$ driven by the temperature gradient. But at large $\chi$ the magnitude is suppressed again. Note that the carrier velocity for the cross term is higher than for the perpendicular, although it similarly decreases with magnetization. 
	
	\section{Idealized transport by a single carrier velocity \label{sec:ideal}}
	
	A generalized coefficient is defined:
	\begin{equation}
		\mathbf{Y} \equiv \int w_Y(v) \mathbf{f}_1 dv
	\end{equation}
	where $\mathbf{f}_1 (v)$ is the anisotropic component of the electron distribution function and $w_Y(v)$ is the velocity weight of that coefficient (representing the curves displayed in figure \ref{fig:lorentz_weights}).  
	
	The coefficients are found from the following integrals:
	\begin{equation}
		Y_\parallel = \int \frac{w_Y S}{\nu} dv \label{eq:Y_parallel_noassum}
	\end{equation}
	\begin{equation}
		Y_\bot + i Y_{\wedge}= \int \frac{w_Y S}{\nu} \frac{1}{1+i\chi(v)} dv \label{eq:Y_perp_noassum}
	\end{equation}
	where $S(v)$ is proportional to the driving term ($\nabla T_e$ or $\mathbf{j}$) and $\nu(v)$ is the collision frequency.

	\subsection{Idealized magnetized Lorentzian coefficient \label{sec:ideal_mag}}
	
	An analytic idealized coefficient is formulated from equations \ref{eq:Y_parallel_noassum} and \ref{eq:Y_perp_noassum} using two assumptions. The first is that the carrier velocity of the electrons is fixed at $v_Y$ with zero spread; for thermal conductivity this is akin to assuming all heat-flux is carried by electrons travelling at $v=2.6v_{th}$. The plasma is also assumed to be Lorentzian ($Z= \infty $) such that e-e collisions are insignificant ($\nu = \nu_{ei}$). This second assumption will be relaxed in section \ref{sec:ideal_ee}.
	
	With these approximations, the Dirac $\delta$ function can be used ($W_Y = Y_\parallel \delta(v-v_Y)$) to obtain:
	\begin{equation}
		Y_\bot = \frac{Y_\parallel}{1+\psi_Y^2} \label{eq:Y_perp_mag}
	\end{equation}
	
	\begin{equation}
		Y_\wedge = \psi_Y Y_\bot \label{eq:Y_wedge_mag}
	\end{equation}
	
	\begin{equation}
		\psi_Y = \chi(v_Y) =  \frac{4}{3\sqrt{\pi}}   \Big(\frac{v_Y}{v_{th}}\Big)^3 \chi \label{eq:Y_psi}
	\end{equation}
	Equation \ref{eq:Y_psi} defines the carrier Hall parameter for this coefficient. At $\psi_Y=1$, $Y_\bot = Y_\parallel/2$. $Y_{\wedge}$ peaks at $\psi_Y=1$ with value $Y_{\wedge}^{peak}= Y_\parallel/2$. Section \ref{sec:ideal_psi} will show this concept to be applicable to the real transport coefficients, with a carrier Hall parameter fit to each process. 
	
	\subsection{Idealized coefficient with electron-electron collisions \label{sec:ideal_ee}}
	
	e-e collisions are added to the idealized single-carrier coefficient by using an effective collision frequency $\nu_{eff}(v) = \nu_{ei}(v) + \nu_{ee}(v)$. The e-i rate is proportional to $n_i Z^2 = n_e Z$ while the e-e rate is proportional to only $n_e$. Therefore $\nu_{ee}/\nu_{ei} = c(v)/Z$, where $c(v)$ is a number that depends on $v$ but not $Z$. The collisional enhancement is therefore:
	
	\begin{equation}
		\frac{\nu_{eff}(v)}{\nu_{ei}(v)} = \frac{Z+c(v)}{Z}
	\end{equation}
	Using the same process as section \ref{sec:ideal_mag}, the parallel coefficient becomes:	
	\begin{equation}
		\frac{Y_\parallel}{Y_\parallel(Z=\infty)} = \frac{Z}{Z+c_Y} \label{eq:Y_par_ee}
	\end{equation}
	Where the Lorentzian value can be found analytically depending on which moment this generalized coefficient represents. 
	
	The perpendicular and wedge coefficients remain the same as equations \ref{eq:Y_perp_mag} and \ref{eq:Y_wedge_mag}, but with the carrier Hall parameter modified:	
	\begin{equation}
		\psi_Y  = \frac{Z}{Z+c_Y} \frac{4}{3\sqrt{\pi}}   \Big(\frac{v_Y}{v_{th}}\Big)^3 \chi  \label{eq:Y_psi_ee}
	\end{equation}
	The same factor of $Z/(Z+c_Y)$ in equations \ref{eq:Y_par_ee} and \ref{eq:Y_psi_ee} suggests that e-e collisions lower the magnitude of the parallel coefficient by the same amount that they reduce the magnetization.
	
	\subsection{The carrier Hall parameter \label{sec:ideal_psi}}
	
	The literature orthodoxy uses the bulk Hall parameter $\chi$ to quantify magnetization. The perpendicular coefficients all equal their parallel counterparts for $\chi=0$, then decrease as $\chi$ increases. However, the consistency between coefficients stops there.
	
	$\chi$ was defined by Braginskii at an electron velocity of $1.10$ times thermal, which is not representative of how each transport process magnetizes. For $Z=1$, $\kappa_{\bot}^c(\chi=1)/\kappa_{\parallel}^c = 0.31$; this falls to $0.048$ for $Z=\infty$. 
	
	The Hall parameter's deficiencies in representing magnetization continue between coefficients. In the Lorentz limit $\gamma_{\bot}^c(\chi=1)/\gamma_{\bot}^c(\chi=0) = 0.037$ compared with $\beta_{\bot}^c(\chi=1)/\beta_{\parallel}^c(\chi=0) = 0.214$. While the current formulation using $\chi$ is suitable for implementation into an MHD code, it is not susceptible to building intuition for how a magnetic field is affecting plasma transport. This behavior was anticipated by the idealized single-carrier model presented in sections \ref{sec:ideal_mag} and \ref{sec:ideal_ee}; different moments of $\mathbf{f}_1$ and different ionizations are magnetized at different rates (equation \ref{eq:Y_psi_ee}). 
	
\begin{figure*}[tb]
		\includegraphics[width=\textwidth]{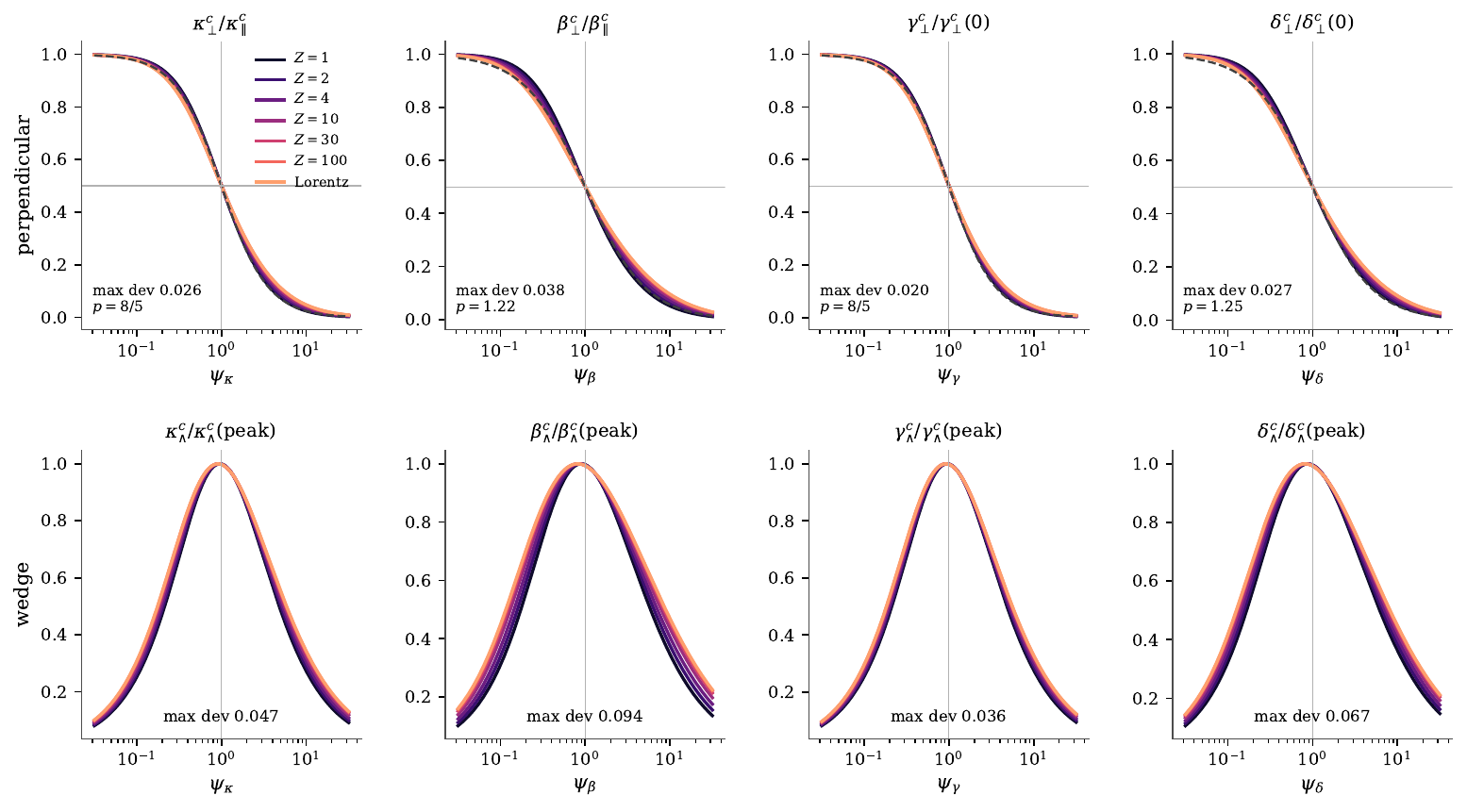}
		\caption{\label{fig:all_coefficients_collapse} The collapse of the magnetization effect for each transport coefficient through the use of a different carrier Hall parameter $\psi$ for each process. The bespoke formulations of $\psi$ are given in table \ref{tab:transport}. Top row: perpendicular ($\bot$) coefficents normalized to the unmagnetized parallel coefficent ($\parallel$) for a range of $Z$. Bottom row: cross ($\wedge$) coefficients normalized to their peak value for a range of $Z$. }
	\end{figure*}

	A carrier Hall parameter ($\psi$) for $\kappa_{\bot}^c$ fixes this issue:
	\begin{equation}
		\psi_{\kappa} = 9.94 \frac{Z+0.17}{Z+5.88} \chi \label{eq:psi_kap}
	\end{equation}
	with the first panel of figure \ref{fig:all_coefficients_collapse} showing the collapse. All $Z$ now approximately follow one magnetization curve. The factor 9.94 in equation \ref{eq:psi_kap} has been chosen to give $\kappa_{\bot}^c/\kappa_{\parallel}^c = 0.5$ at $\psi_{\kappa} = 1$.

	This process has been undertaken to obtain the carrier Hall parameter for all the perpendicular coefficients using the form:
	\begin{equation}
		\psi = A \frac{Z+a}{Z+b} \chi \label{eq:psi_fit}
	\end{equation}
	with the fits given in table \ref{tab:transport} and subsequent collapses in figure \ref{fig:all_coefficients_collapse}. The maximum deviations from the fits have been stated in the figure.
	
\begin{figure}[tb]
		\includegraphics[width=\linewidth]{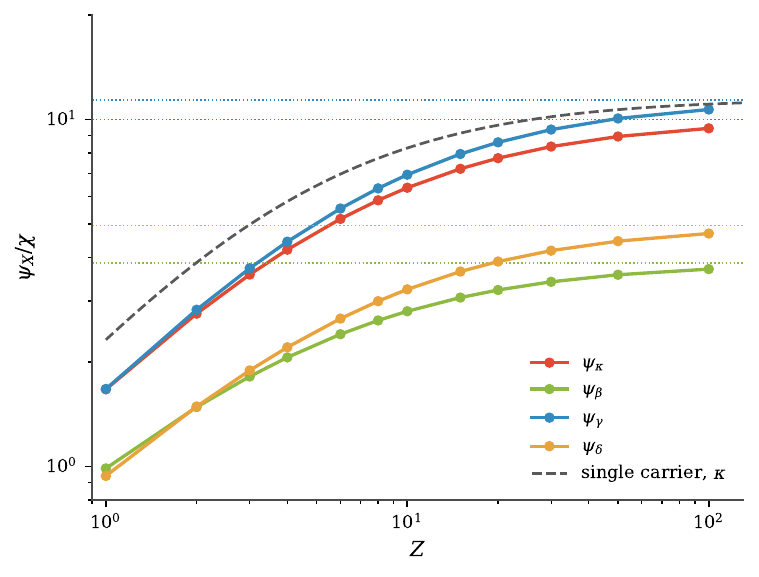}
		\caption{\label{fig:Psi_scales} The carrier Hall parameter ($\psi$) normalized to the traditional Hall parameter ($\chi$) for each coefficient against plasma ionization $Z$. Lorentzian plasmas are more easily magnetized due to there being more high energy electrons in $\mathbf{f}_1$ when e-e collisions are insignificant. $\kappa$ and $\gamma$ coefficients are more easily magnetized due to heat-flux and Nernst advection being carried by higher velocity electrons than the other coefficients.}
	\end{figure}

	Figure \ref{fig:Psi_scales} compares the carrier Hall parameters normalized by $\chi$ for each coefficient. $\kappa_{\bot}^c$ and $\gamma_{\bot}^c$ are more easily magnetized than $\beta_\bot^c$ and $\delta_\bot^c$ due to their dependence on a higher velocity part of the electron distribution. Figure \ref{fig:Psi_scales} also shows that e-e collisions make a plasma harder to magnetize. 
	
	Figure \ref{fig:Psi_scales} also includes $\psi_{\kappa}$ predicted by the idealized single-carrier coefficient, reproducing the general trend but not the magnitude. 
	
	The rate of magnetization is quantified using the form (for thermal conductivity) $\kappa_{\bot}^c / \kappa_{\parallel}^c = 1/(1+\psi_\kappa^p)$. $p=1.50$ is fit to the coefficients, with the other coefficients given in table \ref{tab:transport}. 
	
	The set of cross-coefficients ($\kappa_{\wedge}^c$, $\gamma_{\wedge}^c$, $\beta_{\wedge}^c$ and $\delta_{\wedge}^c$) also require a carrier Hall parameter to collapse. The bottom row of figure \ref{fig:all_coefficients_collapse} shows that sharing the same carrier Hall parameter as their perpendicular counterpart is adequate for collapsing all ionizations onto one curve. Note that the peaks land marginally below $\psi=1$ ($\kappa_\wedge^c$ peaks at $\psi_\kappa\approx0.92$), as the cross fluxes are carried by higher velocity electrons that magnetize more easily.

	\subsection{Idealized transport by a single moving carrier velocity  \label{sec:ideal_moving}}

	The idealized coefficient proposed in section \ref{sec:ideal} assumes transport originates from a population of electrons at the same velocity, with that velocity remaining fixed; the carrier velocity is rigid to e-e collisions and magnetization. 
	
	This section improves that approach for thermal conductivity by assuming a population of carrier electrons at a singular velocity, but with the velocity allowed to change with $Z$ and $\psi$. This is achieved by utilizing the carrier velocity fits presented in appendix \ref{app:carrier}. This is no longer an analytic approach, as it utilizes the measured response of the electron distribution to e-e collisions and magnetization.
	
\begin{figure}[tb]
		\includegraphics[width=\linewidth]{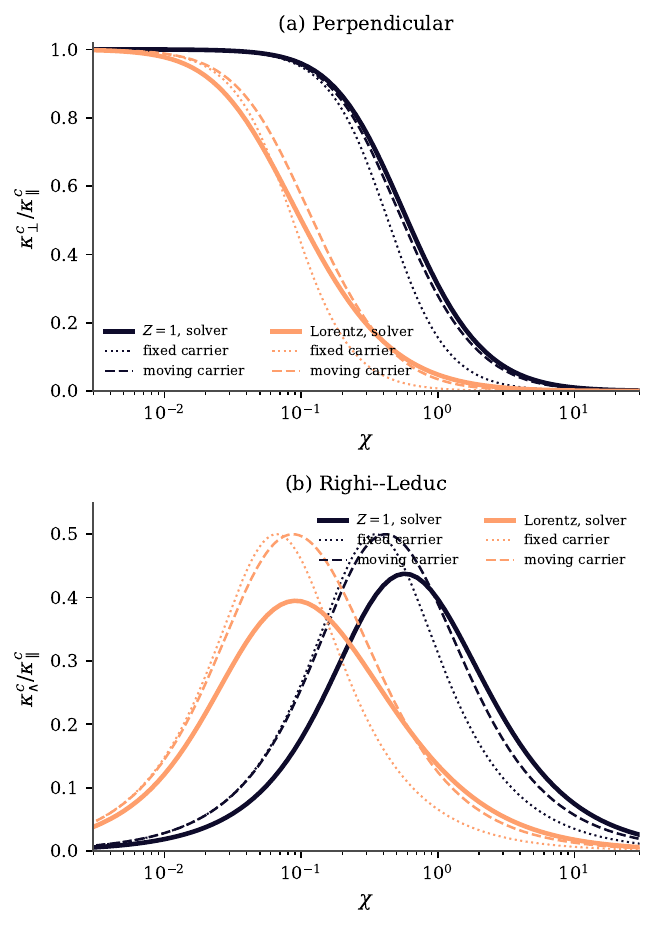}
		\caption{\label{fig:moving_carrier} Perpendicular (a) and Righi-Leduc (b) thermal conductivities against magnetization using different single-carrier models. Solid is the true thermal conductivity. Dotted is the rigid single carrier model outlined in section \ref{sec:ideal}. Dashed is the improved single moving carrier model described in section \ref{sec:ideal_moving}. The remaining difference is due to the real thermal conductivity being carried by a range of electron energies. }
	\end{figure}

	The single-carrier, moving-carrier and fully solved $\kappa_{\bot}^c/\kappa_{\parallel}^c$ are compared in figure \ref{fig:moving_carrier} for $Z=1$ and $Z=\infty$, showing improved agreement with the moving carrier approach. The Righi-Leduc coefficient (figure \ref{fig:moving_carrier}) compares less favorably, albeit still an improvement on the single carrier approach; the true spread of carrier velocities smooths out the curve, lowering the cross-term coefficient. 
	
	\subsection{Limitations of the single-carrier model \label{sec:ideal_limitations}}

	The single-carrier model  predicts how a coefficient magnetizes, but not why the coefficients have the magnitudes and $Z$-dependence that they do. 
	
	A single carrier velocity breaks a fundamental step in calculating the $\nabla T_e$-driven processes: zero current. With just one population, there can be no slow return current to cancel the particle flux (see figure \ref{fig:streams}(a)). 
	The Nernst effect depends on the difference between rotation of the thermal stream and the return current of figure \ref{fig:streams}(c); this cannot be represented by the single carrier model, which erroneously predicts the ratio of Nernst to heat-flow to be constant across $Z$.
	
	The shape of the magnetization curve is set by the spread of carrier velocities, which a single carrier does not have. The fixed carrier gives $p=2$ and a cross-coefficient peak of $Y_\parallel/2$, against $p=1.2$--$1.5$ and peaks of $0.33$--$0.44\,Y_\parallel$ (table \ref{tab:transport}). Letting the carrier move (section \ref{sec:ideal_moving}) recovers $\kappa_\bot^c$, but only by using the measured carrier velocities.
	
	\section{Transport by two carriers \label{sec:two_carrier}}
	
	The single carrier model failed for one reason: transport needs electrons moving in opposite directions at different velocities, with the total flux set by the prescribed current.  This section builds the simplest model with that property. Two carrier velocities turn out to be enough to explain every unmagnetized coefficient to within a few percent. They also explain why each cross coefficient shares the carrier Hall parameter of their perpendicular counterpart.
	
	The kinetic equation is linear in the driving terms ($\mathbf{J}$ and $\nabla T_e$), so $\mathbf{f}_1$ can be split into two components:
	\begin{equation}
		\mathbf{f}_1 = f_{1E}(v)\,\frac{e\mathbf{E}}{T_e} + f_{1T}(v)\,\frac{\nabla T_e}{T_e} , \label{eq:f1_split}
	\end{equation}
	where $f_{1E}$ is the response to an electric field alone and $f_{1T}$ the response to a temperature gradient alone. With no current allowed, $\mathbf{E}$ is set so that the two carry equal and opposite current: $f_{1T}$ is the \emph{thermal stream} and $f_{1E}\,e\mathbf{E}/T_e$ the \emph{return current} (figure \ref{fig:streams}). 
	
	The two carrier populations of this model are \emph{not} separated by whether they are a thermal stream or a current carrier; each population is both. However, the two populations \emph{are} affected differently by an electric field than a temperature gradient.
	
	\subsection{Three properties of the conducting electrons}
\begin{figure}[tb]
		\includegraphics[width=\linewidth]{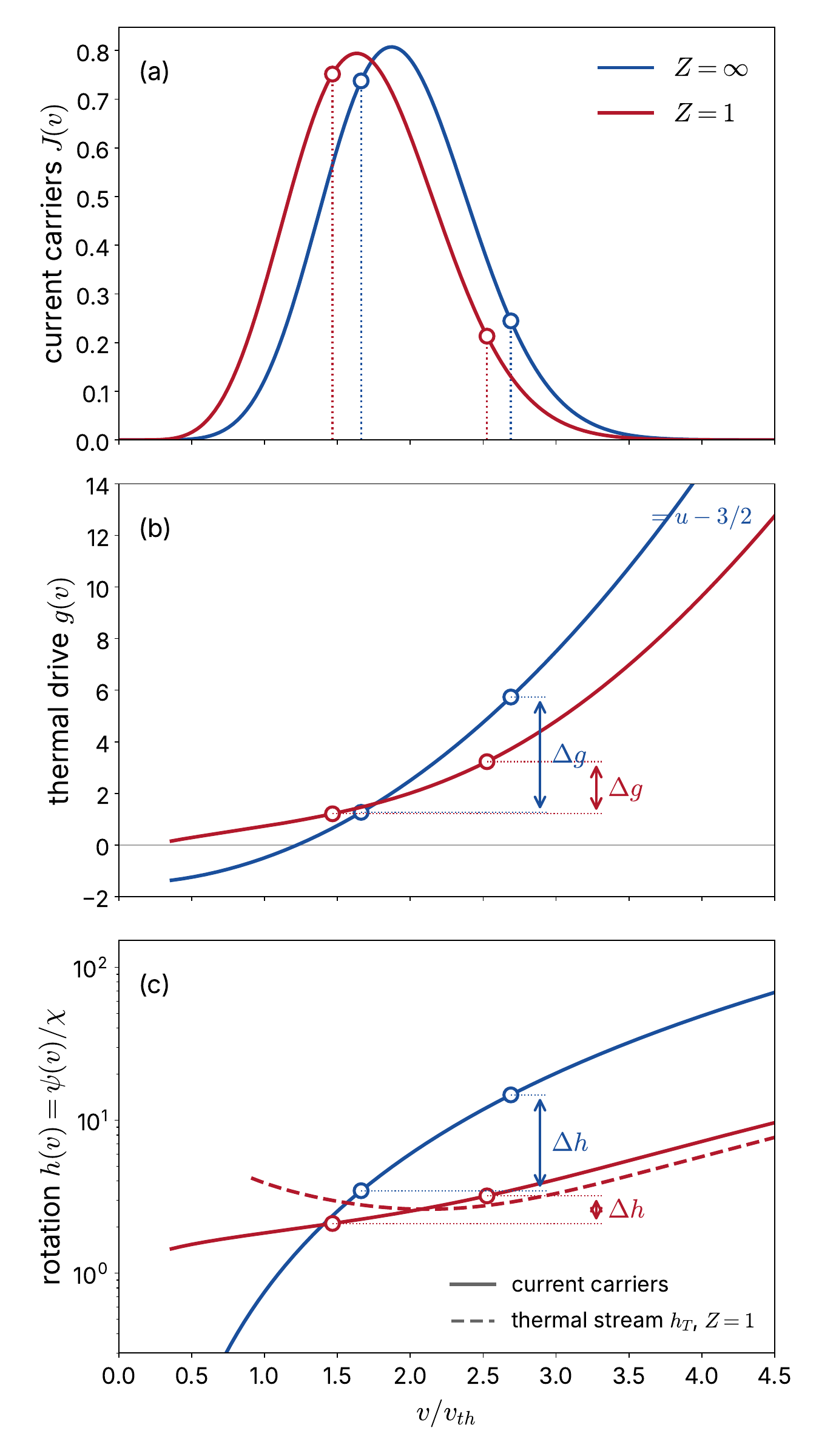}
		\caption{\label{fig:twoc_carriers} (a) Share of the conductivity contributed by each speed, $\mathcal{J}(v)$. This is the same shape as the return current in figure \ref{fig:streams}(a), with the two carriers marked. (b) Thermal drive $g(v)$: how hard the temperature gradient drives each velocity, relative to an electric field. (c) Rotation $h(v)$ of each velocity, normalized to the bulk magnetization $\chi$. Dashed is rotation of the thermal stream at $Z=1$. Arrows mark the drive gap $\Delta g$ and rotation gap $\Delta h$ between the carriers.}
	\end{figure}

	Apply an electric field alone. Every electron responds, but not equally: the response $f_{1E}$ is weighted towards fast electrons, which collide less. Normalizing $W_\alpha = v^3 f_{1E}$ (appendix \ref{app:moments}) to unit area gives the share of the conductivity at each speed,
	\begin{equation}
		\mathcal{J}(v) = \frac{v^3 f_{1E}}{\int v^3 f_{1E}\,dv},
	\end{equation}
	plotted in figure \ref{fig:twoc_carriers}(a) for $Z=1$ and $Z=\infty$. It is also the shape of the return current in figure \ref{fig:streams}(a). Conductivity-weighted averages over $\mathcal{J}$ are written $\langle\cdot\rangle$. The normalization, $\int v^3 f_{1E}\,dv \propto 1/\alpha_\parallel^c$, is the conductivity: how freely the electrons carry current. Each electron drifts for a time $1/\nu(v)$ between collisions, so the conductivity is an average collision time, weighted towards the fast electrons that collide least. e-e collisions at $Z=1$ reduce conductivity to $0.59$ of its Lorentz value.
	
	Three properties of an electron at velocity $v$ give the transport in a plasma  (figure \ref{fig:twoc_carriers}):
	\begin{itemize}
		\item its \emph{energy}, $u = v^2/v_{th}^2$.
		\item its \emph{thermal drive}, $g(v)=f_{1T}/f_{1E}$: how hard a temperature gradient drives electrons at this velocity, relative to an electric field. An electric field pushes every electron with the same force; a temperature gradient instead acts through the imbalance between electrons arriving from the hotter and colder sides, which grows with energy. In the Lorentz limit $g = u-3/2$; fast electrons are driven hardest, and electrons below $u=3/2$ are driven slightly backwards.
		\item its \emph{rotation}, $h(v)$: how far the magnetic field turns the return current at this velocity, per unit $\chi$. In the Lorentz limit this is $h = \chi(v)/\chi$ from equation \ref{eq:chi_v}; fast electrons collide less and are turned further.
	\end{itemize}
	With e-e collisions, $g$ and $h$ are no longer simple functions of $v$; they are read from the solved $\mathbf{f}_1$.
	
	Every unmagnetized coefficient is then a conductivity-weighted average of these differences between streams. These relations are exact at all $Z$:
	\begin{align}
		\beta_\parallel^c &= \langle u\rangle - \tfrac{5}{2}, \label{eq:id_beta}\\
		\kappa_\parallel^c &= \mathrm{Cov}(u,g)/\alpha_\parallel^c, \label{eq:id_kappa}\\
		\gamma_\bot^c(0) &= \mathrm{Cov}(u,h), \label{eq:id_gamma}\\
		\delta_\bot^c(0) &= \alpha_\parallel^c\langle h\rangle - 1, \label{eq:id_delta}
	\end{align}
	where $\mathrm{Cov}(a,b) = \langle ab\rangle - \langle a\rangle\langle b\rangle$ measures how much $a$ and $b$ vary together across $\mathcal{J}(v)$. The thermoelectric coefficient ($\beta_{\parallel}^c$) is the extra energy carried per unit of current, above the enthalpy $5/2$. Thermal conductivity ($\kappa_{\parallel}^c$) is large when the electrons with more energy are also driven harder by the temperature gradient. Nernst ($\gamma_{\bot}^c$) is large when the electrons with more energy are also rotated further.
	
	\subsection{Two carriers}

	Replace $\mathcal{J}(v)$ by just two velocities: a slow carrier $v_1$ with weight $w_1$ and a fast carrier $v_2$ with weight $w_2=1-w_1$. They are chosen so that the two carriers reproduce the total, mean, spread and skewness of the energy distribution of $\mathcal{J}(v)$ (a two-point Gauss quadrature of $\mathcal{J}(v)$), with nothing fitted. In the Lorentz limit they are analytic: $u_{1} = 5-\sqrt{5}$ and $u_{2} = 5+\sqrt{5}$, giving $v_1 = 1.66v_{th}$ and $v_2 = 2.69v_{th}$, with $w_1 = (5+\sqrt{5})/10 = 0.72$. They are marked in figure \ref{fig:twoc_carriers}. 
	
	A comparison of these velocities with the heat-flux moment in figure \ref{fig:ee_on_f1}(b) makes the picture clear: population $2$ are the high energy electrons carrying most of the heat down the gradient, while population $1$ represents the slow return current.
	
	With two values, every covariance becomes a product of differences:
	\begin{equation}
		\mathrm{Cov}(a,b) = w_1 w_2\,\Delta a\,\Delta b,
	\end{equation}
	where $\Delta a = a(v_2) - a(v_1)$. The coefficients then reduce to three \emph{gaps} between the fast and slow carriers: the energy gap $\Delta u$, the drive gap $\Delta g$ and the rotation gap $\Delta h$:
	\begin{align}
		\kappa_\parallel^c &\approx w_1 w_2\,\Delta u\,\Delta g/\alpha_\parallel^c, \label{eq:2c_kappa}\\
		\gamma_\bot^c(0) &\approx w_1 w_2\,\Delta u\,\Delta h, \label{eq:2c_gamma}\\
		\beta_\parallel^c &= w_1u_1 + w_2u_2 - \tfrac{5}{2}, \label{eq:2c_beta}\\
		\delta_\bot^c(0) &\approx \alpha_\parallel^c\,(w_1h_1 + w_2h_2) - 1. \label{eq:2c_delta}
	\end{align}
		The first is exact, as the two carriers reproduce the mean energy by construction; the others are within $1.7\%$, $3.2\%$ and $2.0\%$ of the full solution for all $Z$. In the Lorentz limit the thermal conductivity is exact: $w_1w_2 = 1/5$ and $\Delta u = \Delta g = 2\sqrt{5}$ give $\kappa_\parallel^c = 4/\alpha_\parallel^c = 128/3\pi$.
	
	Each equation has a simple picture:
	
	\textbf{Thermal conductivity.} The temperature gradient drives the fast carrier harder than the slow one ($\Delta g$). The return current removes the average drive. The fast carrier streams down the gradient and the slow carrier streams back in equal numbers, but they carry different energies ($\Delta u$). The net energy flux is the heat-flux.
	
	\textbf{Nernst.} The return current is deflected in the opposite direction to the electrons carrying flux. If the energies were the same, these fluxes would cancel. The faster electrons are deflected further ($\Delta h$) so there is a net sideways flux. The $E_y$ that balances particle flux (a transverse return current) is the source of Nernst.
		
	\textbf{$\beta_{\parallel}^c$} The energy carried by the carriers, minus the energy carried by the bulk current.
	
	\textbf{$\delta_{\bot}^c(0)$} The current as a whole rotates by the average rotation.

	\subsection{Two carriers with e-e collisions ($Z<\infty$)}
	
\begin{figure}[tb]
		\includegraphics[width=\linewidth]{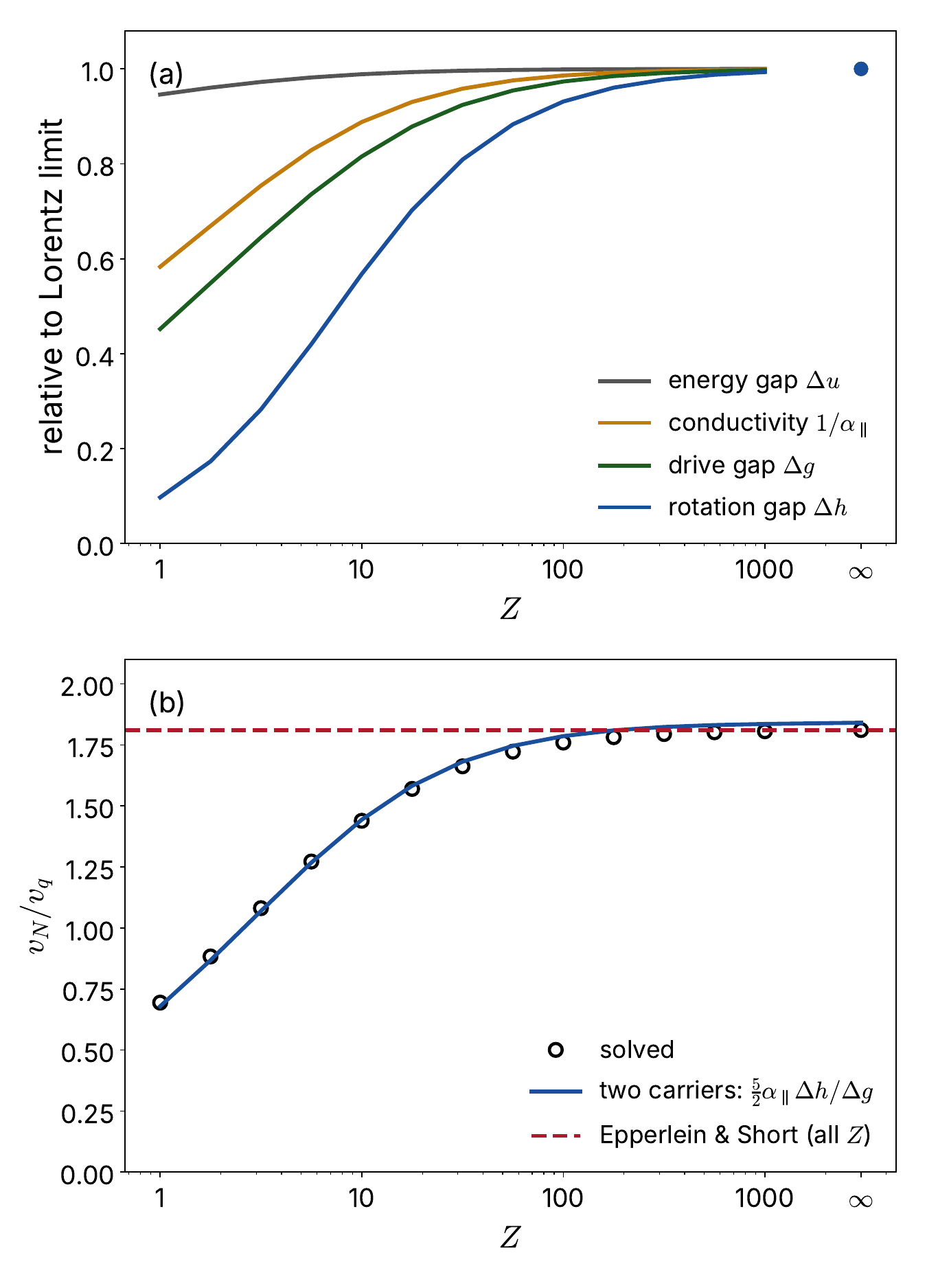}
		\caption{\label{fig:twoc_gaps} (a) The gaps between the two carriers, relative to the Lorentz limit. e-e collisions barely change the energy gap, halve the drive gap and remove nine tenths of the rotation gap. (b) Nernst velocity relative to heat-flow velocity: full solution (points), two carriers (equation \ref{eq:vN_vq}) and the Epperlein \& Short approximation.}
	\end{figure}

	Figure \ref{fig:twoc_gaps}(a) shows how each gap changes from the Lorentz limit to $Z=1$. The carriers move little, from $1.66 \to 1.47v_{th}$ and $2.69\to 2.53v_{th}$, so the energy gap $\Delta u$ falls by only $5\%$. The drive gap falls by 50\%. The rotation gap $\Delta h$ falls by a factor of 10. These differences are explained in the rest of this section.

	First, e-e collisions raise the overall collision rate, with the conductivity falling to $0.59$ of its Lorentz value at $Z=1$. e-e collisions also change how fast and slow carriers respond to $\nabla T_e$. The fast carrier shares the drive it receives from $\nabla T_e$ with the slow electrons, closing the drive gap. Momentum is passed from the fast electrons to the slow ones: at $Z=1$ the fast carrier's drive is $3.2$, against $4.9$ for a Lorentz plasma at the same speed, while the slow carrier's drive rises from $0.7$ to $1.2$.
	
	The drive is a ratio of two responses of the same electrons, so a higher collision rate cancels out of it; the rotation is the turn between collisions, so the extra rate enters in full. The rotation is affected more strongly because the fast carrier barely collides with ions, so e-e collisions are a large addition to its collision rate. Its rotation falls by a factor of $3.8$. The slow carrier already collides often with ions, so its rotation falls by only $12\%$. Heat-flow loses 50\% of its gap ($\Delta g$) and Nernst loses 90\% ($\Delta h$), which is why e-e collisions suppress Nernst far more than heat-flow.
	
	The Nernst velocity relative to the heat-flow velocity is $v_N/v_q = (5/2)\gamma_\bot^c/\kappa_\bot^c$. In the two-carrier model the factor $w_1w_2\Delta u$ cancels, leaving
	\begin{equation}
		\frac{v_N}{v_q} = \frac{5}{2}\,\alpha_\parallel^c\,\frac{\Delta h}{\Delta g}. \label{eq:vN_vq}
	\end{equation}
	The ratio of Nernst to heat-flow is the rotation gap over the drive gap. This gives $1.84$ in the Lorentz limit and $0.70$ at $Z=1$, against the full solution's $1.81$ and $0.69$ (figure \ref{fig:twoc_gaps}(b)).
	
	The simplified collision operator of Epperlein \& Short\cite{epperleinPracticalNonlocalModel1991} is a uniform rescaling of the collision rate. By design, it matches the thermal conductivity. However, $\Delta h/\Delta g$ keeps its Lorentz value and $v_N/v_q = 1.81$ at every $Z$ (figure \ref{fig:twoc_gaps}(b)). It is not a useful collision model for any coefficient except thermal conductivity (see appendix \ref{sec:ee_colls}).
	
	\subsection{Magnetization in the two carrier model}
	The single-carrier model assumed that each perpendicular and cross coefficient share one carrier Hall parameter. The two-carrier model shows this is a consequence of the zero-current condition for the $\nabla T_e$-driven terms. For the Nernst coefficients:
	\begin{align}
		\gamma_\bot^c &= \frac{\gamma_\bot^c(0)}{1+\psi^2}, &
		\gamma_\wedge^c &= \psi\,\gamma_\bot^c, &
		\psi &= (w_1 h_2 + w_2 h_1)\,\chi. \label{eq:2c_mag}
	\end{align}
	This is the single-carrier magnetization law (equations \ref{eq:Y_perp_mag} and \ref{eq:Y_wedge_mag}), now with a carrier Hall parameter that follows from the two rotations. In the Lorentz limit $\psi = 11.6\chi$, against the fitted $\psi_\gamma = 11.36\chi$ (table \ref{tab:transport}). The same holds for $\kappa^c$ in the Lorentz limit, where the thermal stream rotates with the current carriers. 
	
	The resistivity ($\alpha^c$, $\delta^c$) has no zero-current subtraction, with the two carriers conducting in parallel. Each carrier population is turned by its own rotation, so the perpendicular resistivity is a sum of two single-carrier curves. Nonetheless, the two components of $\delta^c$ are found to follow the same carrier Hall parameter (figure \ref{fig:all_coefficients_collapse}).
	
	At weak field the thermal cross terms are:
	\begin{align}
		\lim_{\chi\to0}\kappa_\wedge^c/\chi &= \big[\mathrm{Cov}(u,g\,h_T) - \langle g\rangle\,\gamma_\bot^c(0)\big]/\alpha_\parallel^c, \label{eq:id_kwedge}\\
		\lim_{\chi\to0}\gamma_\wedge^c/\chi &= \mathrm{Cov}(u,h^{(2)}) - \langle h\rangle\,\gamma_\bot^c(0). \label{eq:id_gwedge}
	\end{align}
	Here $h_T$ is the rotation of the thermal stream (dashed in figure \ref{fig:twoc_carriers}(c)) and $h^{(2)}$ is the return current rotated a second time. e-e collisions tie each velocity to the flow of the electrons around it, so the thermal stream turns by a different amount, $h_T$. Without e-e collisions these are $h$ and $h^2$. Both cross terms therefore depend on how differently the two streams are rotated.
	
	The two-carrier model still gives the exponent on $\psi$ to be 2, which decreases transport too quickly with magnetization. As in section \ref{sec:ideal_moving}, the fix is to let the carriers move: at each $\chi$ the two carriers are re-chosen from the electrons not yet magnetized. Figure \ref{fig:twoc_mag} shows that the moving pair reproduces $\kappa_\bot^c$ and $\beta_\wedge^c$ to within $6\%$ in the Lorentz limit over four decades of $\chi$, and to within $\sim20\%$ at $Z=1$. At strong field both carriers settle at $0.77v_{th}$ and $1.85v_{th}$ for every $Z$: the electrons that remain unmagnetized drift together, and only collisions with ions can stop them.
	
	\subsection{Limitations of the two carrier model}
	The two-carrier model is used here as a description of the coefficients; $\Delta u$, $\Delta g$ and $\Delta h$ are taken from the solved distributions, rather than predicting the gaps. 
	
	Could the gaps be predicted instead? The obvious attempt keeps the two $Z=\infty$ carrier populations and adds e-e collisions. This misses the $Z$-dependence by up to $37\%$ at $Z=1$. The reason is momentum. The two carrier populations are faster, and so less collisional, than the bulk: the momentum e-e collisions take from them must pass to slow electrons before the ions can remove it, and the two carriers include no such population. Section \ref{sec:shapes} fixes this issue by including a slow third population.
	
\begin{figure}[tb]
		\includegraphics[width=\linewidth]{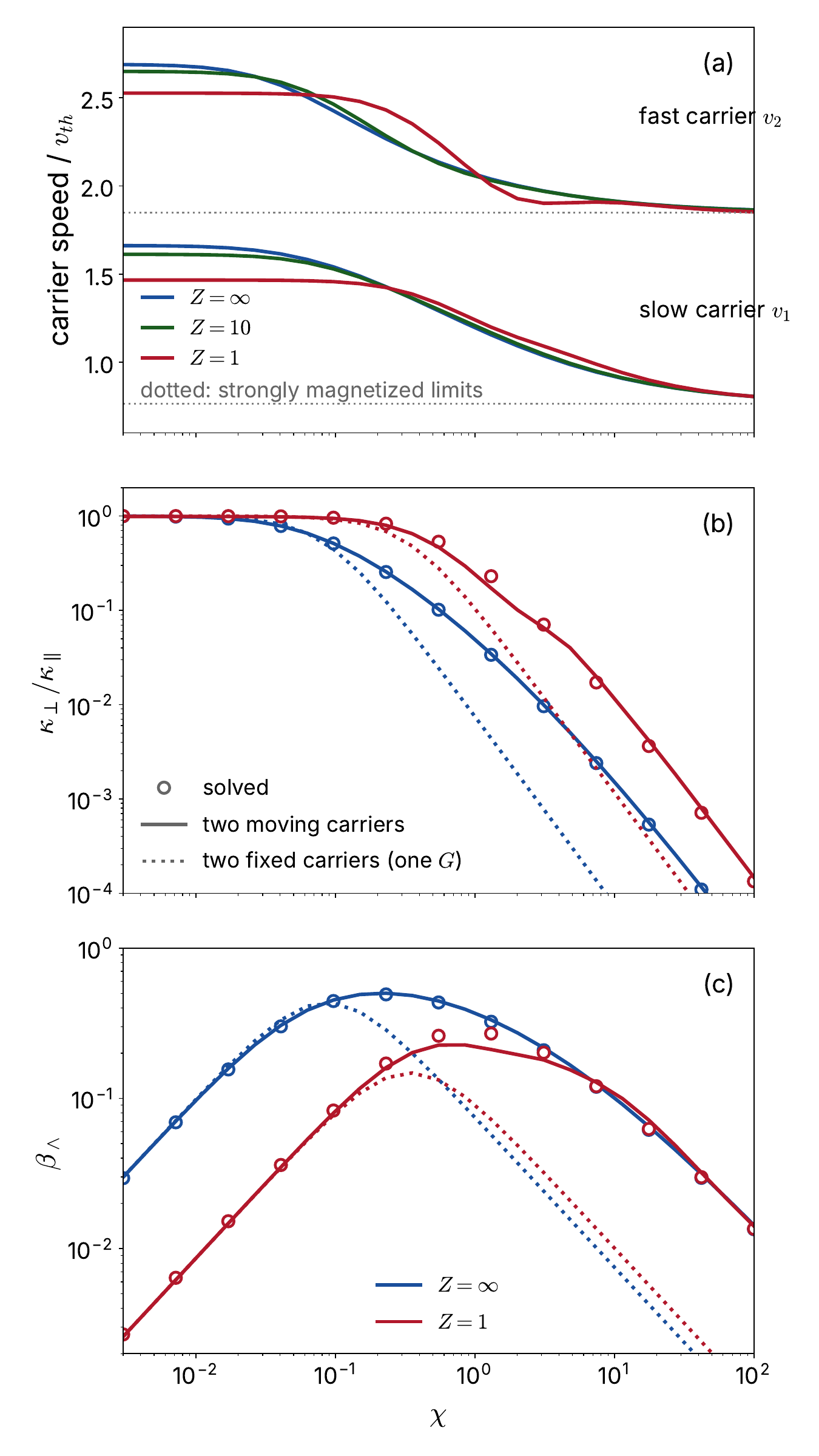}
		\caption{\label{fig:twoc_mag} (a) Speeds of the two moving carriers against $\chi$; dotted: the strongly magnetized limit. (b) $\kappa_\bot^c/\kappa_\parallel^c$ and (c) $\beta_\wedge^c$: full solution (points), two moving carriers (solid) and two fixed carriers with one carrier Hall parameter (dotted).}
	\end{figure}

	Additionally, the two-carrier model magnetizes the coefficients as $1/(1+\psi^2)$, while the true exponent is $1.5$. As with the single-carrier coefficient (section \ref{sec:ideal_moving}), allowing the carrier velocity to decrease with magnetization fixes this issue (figure \ref{fig:twoc_mag}). However, this introduces extra parameters taken from the coefficients, rather than predicting them.
	
	\section{Two carriers and a slow population \label{sec:shapes}}
	
	The two-carrier model uses the solved distributions associated with the coefficients to get the energy, drive and rotation differences between the two carrier populations. To predict them, the momentum that e-e collisions take from the two populations (both of which are above thermal velocity) needs somewhere to go. e-e collisions conserve momentum, so they cannot remove momentum on their own; only ions can. This mostly happens through electrons slower than the two carrier populations.
	
	The missing piece is a \textbf{drifting Maxwellian}, $v f_0$: every electron moving with the same velocity.  e-e collisions take momentum from the carriers and put it into this shared drift, where the ions remove it. Adding it brings every coefficient's $Z$-dependence to within $\sim10\%$, with nothing fitted. 
	
	The drift is why conductivity falls at low $Z$. A shared drift is stopped by the ions mostly through its slow electrons, whose collision rate rises as $v^{-3}$, so current moved into the drift meets more resistance than current carried by fast electrons. As $Z$ grows, e-e collisions weaken relative to e-i collisions. Less momentum is handed over, and the drift's share of the current falls roughly as $1/Z$ (from $29\%$ of the current at $Z=1$ to $6\%$ at $Z=5$), vanishing in the Lorentz limit.
	
\begin{figure}[tb]
		\includegraphics[width=\linewidth]{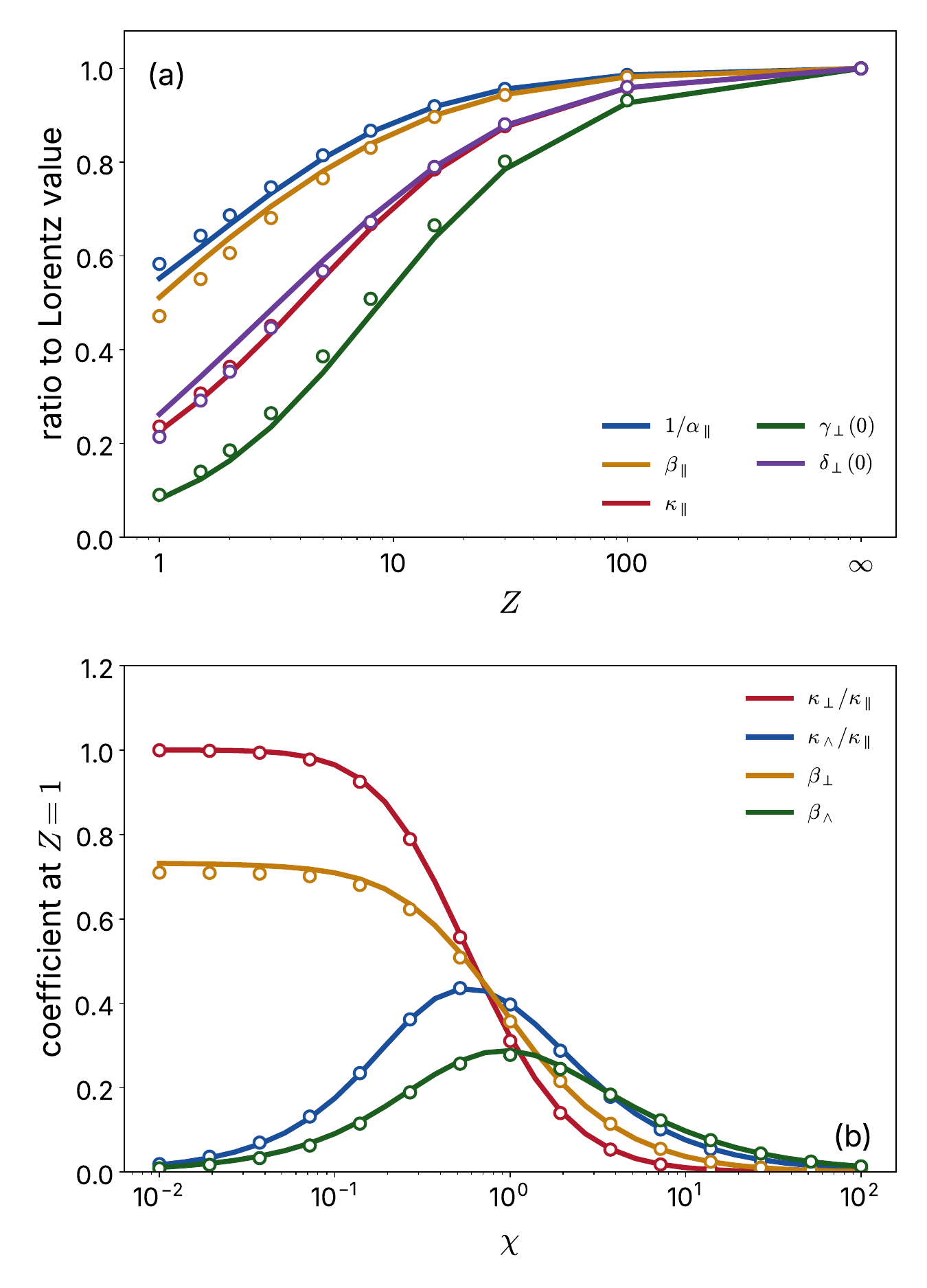}
		\caption{\label{fig:threepop} (a) Unmagnetized coefficients relative to their Lorentz values. Full solution (points) and the three-population model (lines). (b) Coefficients at $Z=1$ against $\chi$: full solution (points) and the three-population model (lines). Nothing is fitted.}
	\end{figure}

	Figure \ref{fig:threepop}(a) compares the three population model with the full solution across $Z$. Nothing is fitted. Every unmagnetized coefficient follows the full solution, with the largest errors at low $Z$: $5\%$ in the conductivity, $8\%$ in $\beta_\parallel^c$, $4\%$ in $\kappa_\parallel^c$, $12\%$ in $\gamma_\bot^c(0)$ and $22\%$ in $\delta_\bot^c(0)$.
	
	To magnetize the model, the two carrier populations are rotated by the field using $\chi(v)$ of equation \ref{eq:chi_v}. The shared drift is slow and needs no rotation. The fast electrons are turned furthest, so the flux that remains moves to slower electrons on its own: at $Z=1$ the heat-flow carrier slows from $2.38v_{th}$ to $1.86v_{th}$ at strong field, against $1.88v_{th}$ in the full solution. Figure \ref{fig:threepop}(b) shows the result at $Z=1$. $\kappa_\bot^c$, $\kappa_\wedge^c$ and $\beta_\bot^c$ are within $8\%$ at every $\chi$, and in the Lorentz limit the model is exact.
	
	The exception is Nernst at weak field and low $Z$, seen as a $16\%$ error in $\beta_\wedge^c=\chi\gamma_\bot^c$ near $\chi=0.03$. 
	
	The approach here is similar to the route taken by Braginskii\cite{braginskiiTransportProcessesPlasma1965}, with $\mathbf{f}_1$ written as a sum of a few fixed functions of speed. Braginskii used Laguerre polynomials times a Maxwellian, up to third order. What differs here is the choice of the remaining functions. The two carriers are the exact Lorentz responses ($v^4f_0$ and $v^6f_0$), which a polynomial expansion only approaches with many terms. This is why Braginskii's coefficients are least accurate at strong magnetization, later corrected by Epperlein \& Haines by solving the kinetic equation in full\cite{epperleinPlasmaTransportCoefficients1986}. 
	
	As well as a drifting Maxwellian ($vf_0$), Braginskii employed a drift rising with energy ($v^3f_0$). Adding this fourth population to the model here improves accuracy to within 1\% of the coefficients. As this paper's intention is to explain the coefficients in the carrier framework, rather than to reproduce the coefficients exactly, this has been left out of the physical picture.
	
		\begin{table}[tb]
		\centering
		\small
		\setlength{\tabcolsep}{5pt}
		\renewcommand{\arraystretch}{1.25}
		\caption{\label{tab:transport} Carrier Hall parameter for each coefficient, $\psi_X = A\,\frac{Z+a}{Z+b}\,\chi$, defined so that $\psi_X=1$ halves the coefficient magnitude. The perpendicular coefficients then follow $X_\bot/X_\parallel \simeq 1/(1+\psi_X^{\,p})$; a single carrier gives $p=2$. The cross coefficients share $\psi_X$ and peak at the height given, as a fraction of the unmagnetized coefficient over $Z=1$--$\infty$; the single carrier model gives $0.5$.}
		\begin{tabular}{l c c c c c c}
			\toprule
			& $A$ & $a$ & $b$ & $\dfrac{\psi(Z{=}1)}{\psi(Z{=}\infty)}$ & $p$ & $\wedge$ peak \\
			\midrule
			$\psi_\kappa$  & 9.94  & 0.17 & 5.88 & 0.167 & 1.50 & 0.40--0.44 \\
			$\psi_\beta$   & 3.83  & 0.33 & 4.07 & 0.256 & 1.22 & 0.33--0.39 \\
			$\psi_\gamma$  & 11.36 & 0.13 & 6.58 & 0.147 & 1.52 & 0.40--0.44 \\
			$\psi_\delta$  & 4.93  & 0.28 & 5.62 & 0.190 & 1.25 & 0.33--0.39 \\
			\bottomrule
		\end{tabular}
	\end{table}
	

	
	\section{Conclusion \label{sec:conc}}
	
	Classical transport coefficients have been explained by the electrons that carry them. 
	
	A model with three populations (two opposing carriers and a slow drifting Maxwellian) predicts the coefficients within about $10$--$20\%$ at $Z=1$, with nothing fitted. The two carriers represent a fast population streaming down the gradient and a slow return current; differences in energy, drive and rotation of the opposing streams sets the coefficients. The third slow population allows for reasonable momentum exchange with the ions.
	
	This contrasts with the coefficient fits of Sadler-Walsh-Li (SWL)\cite{sadlerSymmetricSetTransport2021}, which reproduce the coefficients with 136 constants, none of which has a physical reading. The populations here have no free constants, with each being a fixed function of speed. Only how much of each population is present is solved for. This work is not intended to replace the established fits, but shows the physics of what gives them their values.
	
	In particular, this work aids in answering the simple question: how magnetized is a transport process in any given plasma? As $\kappa_{\bot}^c$ (thermal conduction) is carried by higher velocity electrons than $\delta_{\bot}^c$ (heat carried by current), $\kappa_{\bot}^c$ is magnetized more easily. e-e collisions raise the carriers' collision rate (and slow them slightly), making low $Z$ plasmas more difficult to magnetize than high-$Z$ plasmas. The Hall parameter $\chi$ is therefore not a useful metric for quantifying plasma transport magnetization. For this purpose a carrier Hall parameter $\psi$ is fitted for each coefficient, defined such that $\psi=1$ means 50\% transport reduction independent of coefficient or plasma species.
	
	The two-carrier picture also offers a view of kinetic transport, the subject of the next paper in this series\cite{walsh_paper1}. When the temperature varies over distances comparable to the electron mean free path, the two carriers themselves are unchanged: who carries the current is set by the collisions alone. What changes is how a temperature gradient drives the fast carrier. Fast electrons travel far before sharing their energy, so at any point their population reflects the temperature averaged over that distance rather than the local temperature; this is the tail depletion responsible for kinetic suppression of heat-flow. The slow carrier remains local. Kinetic heat-flow suppression can therefore be read as the fast carrier's drive being smoothed over several delocalization lengths, shrinking the drive gap while leaving the energy gap untouched.
	
	\begin{acknowledgments}
		This work was performed under the auspices of the U.S. Department of Energy by Lawrence Livermore National Laboratory under Contract DE-AC52-07NA27344. 
		
		This document was prepared as an account of work sponsored by an agency of the United States government. Neither the United States government nor Lawrence Livermore National Security, LLC, nor any of their employees makes any warranty, expressed or implied, or assumes any legal liability or responsibility for the accuracy, completeness, or usefulness of any information, apparatus, product, or process disclosed, or represents that its use would not infringe privately owned rights. Reference herein to any specific commercial product, process, or service by trade name, trademark, manufacturer, or otherwise does not necessarily constitute or imply its endorsement, recommendation, or favoring by the United States government or Lawrence Livermore National Security, LLC. The views and opinions of authors expressed herein do not necessarily state or reflect those of the United States government or Lawrence Livermore National Security, LLC, and shall not be used for advertising or product endorsement purposes.
		
		The data that support the findings of this study are available from the corresponding author upon reasonable request. 
	\end{acknowledgments}

	\appendix
	
	\section{Moments \label{app:moments}}

	The linearized response is driven by either a current or a temperature
	gradient.  The current depends on $f_1$ as:
	\begin{equation}
		J[f_1] = -\frac{4\pi e}{3}\int_0^\infty v^3 f_1\,\mathrm{d}v
	\end{equation}

	The response per unit current, $f_{1J}$, and the response at zero current,
	$f_1|_{J=0}$:
	\begin{equation}
		f_{1J} = \frac{f_{1E}}{J[f_{1E}]},
		\qquad
		f_1\big|_{J=0} = f_{1T} - J[f_{1T}]\,f_{1J},
		\label{eq:responses}
	\end{equation}
	where $f_{1E}$ and $f_{1T}$ are the responses to an applied electric field and to
	$\nabla T_e/T_e$. 
	
	Each coefficient is then a single velocity moment of one of these responses:
	\begin{align}
		W_\alpha(v)  &= v^3 f_{1E},
		\label{eq:Walpha}\\[2pt]
		W_\beta(v)   &= v^3 f_{1J}\left[v^2 - \frac{5T_e}{m}\right],
		\label{eq:Wbeta}\\[2pt]
		W_\kappa(v)  &= v^5 f_1\big|_{J=0},
		\label{eq:Wkappa}\\[2pt]
		W_\gamma(v)  &= v^5 \frac{\partial f_{1J,z}}{\partial\chi},
		\label{eq:Wgamma}\\[2pt]
		W_\delta(v)  &= v^3 f_{1J}\left[\alpha_\parallel^c\left(\frac{v}{v_B}\right)^3 - 1\right],
	\end{align}
	with $v_B = 1.10v_{th}$ coming from Braginskii's definition of the collision time $\tau_B$. The integrals give the coefficients themselves:
	\begin{align}
		\alpha_\parallel &= \frac{\tau_B}{J[f_{1E}]}
		= -\frac{3\tau_B}{4\pi e}\left[\int_0^\infty W_\alpha\,\mathrm{d}v\right]^{-1}
		\;\xrightarrow[\ Z\to\infty\ ]{}\; \frac{3\pi}{32},
		\label{eq:alpha}\\[2pt]
		\beta_\parallel  &= \frac{m}{2T_e}\int_0^\infty W_\beta\,\mathrm{d}v
		\;\xrightarrow[\ ]{}\; \frac{3}{2},
		\label{eq:beta}\\[2pt]
		\kappa_\parallel &= -\frac{2\pi m}{3\,T_e^2\tau_B}\int_0^\infty W_\kappa\,\mathrm{d}v
		\;\xrightarrow[\ ]{}\; \frac{128}{3\pi},
		\label{eq:kappa}\\[2pt]
		\gamma_\perp(\chi\!\to\!0) &= -\frac{m}{2T_e}\int_0^\infty W_\gamma\,\mathrm{d}v
		\;\xrightarrow[\ ]{}\; \frac{315}{32},
		\label{eq:gamma}\\[2pt]
		\delta_\perp(\chi\!\to\!0) &= \int_0^\infty W_\delta\,\mathrm{d}v
		\;\xrightarrow[\ ]{}\; \frac{945\pi}{1536}-1 .
		\label{eq:delta}
	\end{align}

	\section{Carrier velocities \label{app:carrier}}
	
	%
	
	The mean speed of the electrons carrying each coefficient is reproduced, for every
	$Z$ and every magnetization, by
	\begin{equation}
		\frac{\langle v\rangle_X}{v_{th}}
		= v_\infty + \frac{v_0(Z)-v_\infty}{1+\big(C\,\psi_X\big)^{p}},
		\qquad
		v_0(Z) = v_L\,\frac{Z+a}{Z+b},
		\label{eq:carrier}
	\end{equation}
	where $v_0$ is the unmagnetized carrier, $v_\infty$ the strongly magnetized one, and
	$C\psi_X=1$ marks the field at which the carrier has slowed half way between them.
	Setting $\psi_X=0$ recovers the carrier of the corresponding parallel coefficient.
	
	The constants are:
	\begin{alignat}{2}
		&\kappa^c_\perp\!: &\quad \frac{\langle v\rangle}{v_{th}} &= 1.848+\frac{v_0-1.848}{1+(0.615\,\psi_\kappa)^{1.18}}, \label{eq:vkappa}\\
		&\text{heat-flow} & v_0 &= 2.482\,\frac{Z+2.91}{Z+3.07}, \quad 1.1\% \notag\\[4pt]
		&\kappa^c_\wedge\!: & \frac{\langle v\rangle}{v_{th}} &= 2.165+\frac{v_0-2.165}{1+(0.942\,\psi_\kappa)^{1.39}}, \label{eq:vkappaw}\\
		&\text{Righi-Leduc} & v_0 &= 2.720\,\frac{Z+3.32}{Z+3.60}, \quad 0.7\% \notag\\[4pt]
		&\beta^c_\perp\!: & \frac{\langle v\rangle}{v_{th}} &= 1.598+\frac{v_0-1.598}{1+(0.633\,\psi_\beta)^{0.90}}, \label{eq:vbeta}\\
		&\text{thermoelectric} & v_0 &= 2.325\,\frac{Z+1.08}{Z+1.37}, \quad 1.5\% \notag\\[4pt]
		&\beta^c_\wedge\!: & \frac{\langle v\rangle}{v_{th}} &= 1.210+\frac{v_0-1.210}{1+(0.516\,\psi_\beta)^{0.94}}, \label{eq:vbetaw}\\
		& & v_0 &= 2.690\,\frac{Z+1.89}{Z+2.26}, \quad 2.4\% \notag\\[4pt]
		&\gamma^c_\perp\!: & \frac{\langle v\rangle}{v_{th}} &= 1.840+\frac{v_0-1.840}{1+(0.482\,\psi_\gamma)^{1.28}}, \label{eq:vgamma}\\
		&\text{Nernst} & v_0 &= 2.538\,\frac{Z+2.69}{Z+2.91}, \quad 1.5\% \notag\\[4pt]
		&\gamma^c_\wedge\!: & \frac{\langle v\rangle}{v_{th}} &= 2.173+\frac{v_0-2.173}{1+(0.617\,\psi_\gamma)^{1.32}}, \label{eq:vgammaw}\\
		&\text{cross-Nernst} & v_0 &= 2.767\,\frac{Z+3.20}{Z+3.54}, \quad 1.9\% \notag\\[4pt]
		&\delta^c_\perp\!: & \frac{\langle v\rangle}{v_{th}} &= 1.263+\frac{v_0-1.263}{1+(0.621\,\psi_\delta)^{0.96}}, \label{eq:vdelta}\\
		&\text{Hall} & v_0 &= 2.394\,\frac{Z+2.91}{Z+3.32}, \quad 2.9\% \notag\\[4pt]
		&\delta^c_\wedge\!: & \frac{\langle v\rangle}{v_{th}} &= 1.804+\frac{v_0-1.804}{1+(0.618\,\psi_\delta)^{1.22}}, \label{eq:vdeltaw}\\
		&\text{cross-Hall} & v_0 &= 2.597\,\frac{Z+2.00}{Z+3.13}, \quad 5.1\% \notag
	\end{alignat}
	
	The parallel conductivity $1/\alpha^c_\parallel$ does not depend on the field and needs only the first factor,
	\begin{equation}
		\frac{\langle v\rangle_{1/\alpha^c_\parallel}}{v_{th}}
		= 1.938\,\frac{Z+1.73}{Z+2.07}
		\qquad (0.2\%).
		\label{eq:valpha}
	\end{equation}
	
	%
	%
	%
	%

	\section{Importance of e-e collision operator \label{sec:ee_colls}}
	This section demonstrates why a full e-e collision operator is required for plasma transport.
	
	Three treatments of the e-e collisions are considered in the classical limit, using $Z=1$ (where e-e collisions are most important). First, completely neglecting e-e collisions (equivalent to the Lorentz limit). Second, using the Epperlein \& Short (ES) approximation, which scales the e-i collision frequency uniformly across the electron distribution by $(Z+4.2)/(Z+0.24)$\cite{epperleinPracticalNonlocalModel1991}. Finally, using the full e-e collision operator\cite{tzoufrasVlasovFokkerPlanck2011}, which is velocity dependent.
	
\begin{figure}[tb]
		\includegraphics[width=\linewidth]{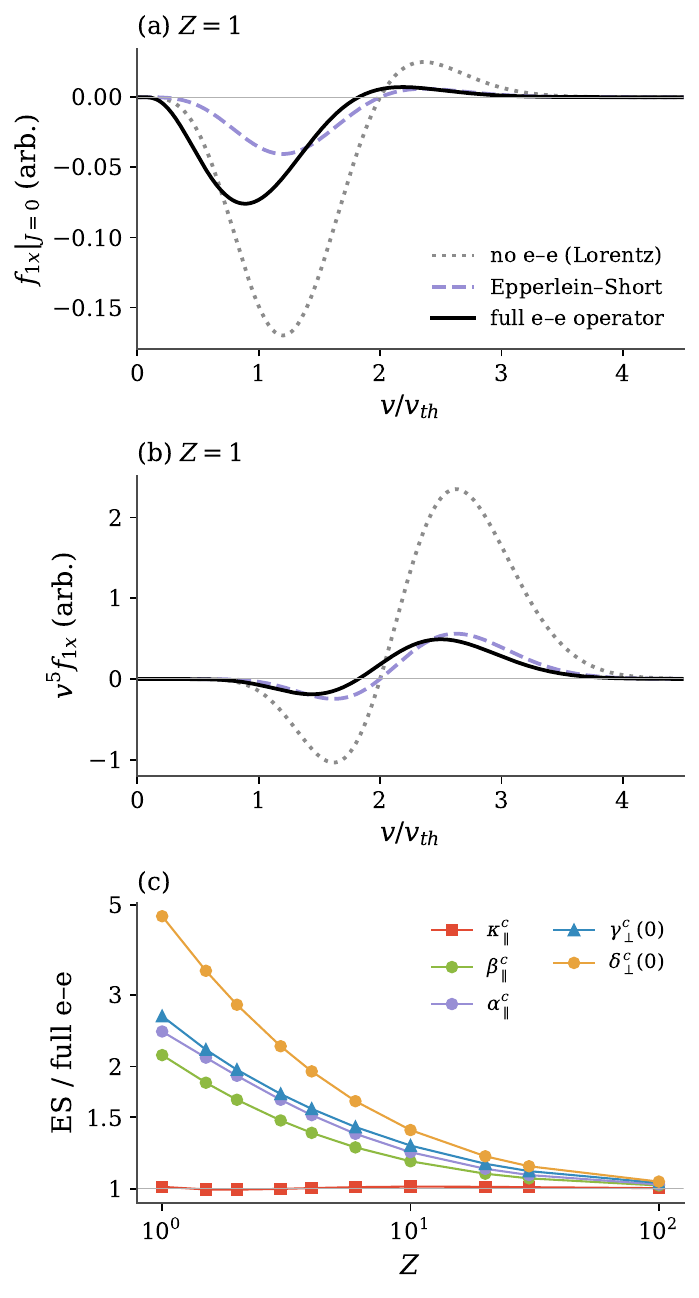}
		\caption{\label{fig:ee_vs_ei} (a) $\mathbf{f}_1$ electron velocity distribution and (b) heat-flow velocity moment for a Z=1 plasma in the classical limit. Shown are 3 cases: no e-e collisions; Epperlein \& Short (ES) approximation for e-e collisions\cite{epperleinPracticalNonlocalModel1991}; full e-e collision operator. The approximation reduces the magnitude of the heat-flow velocity moment, which allows it to reproduce the classical thermal conductivity; this is by design. (c) the ratio of the coefficients calculated using the ES approximation to the true coefficient (using the full operator).  While the ES approximation reproduces thermal conductivity well, it overpredicts the other coefficients at low $Z$. }
	\end{figure}

	Figure \ref{fig:ee_vs_ei}(a) shows the solved $\mathbf{f}_1$ against $v/v_{th}$ for the 3 cases. Figure \ref{fig:ee_vs_ei}(b) then shows how that affects the heat-flow moment ($v^5 f_1$). The main impact of e-e collisions is to reduce the magnitude of the $v^5 f_1$ moment, thereby reducing the thermal conductivity to $\kappa(Z=1)/\kappa(Z=\infty) = 0.236$. Figure \ref{fig:ee_vs_ei}(c) shows that the ES approximation reproduces the $Z$-dependence of the conductivity. Note that this is by design: $\kappa^c(Z=1)/\kappa^c(Z=\infty) \approx (Z+0.24)/(Z+4.2)$.
	
	What the Epperlein \& Short approximation does not account for is the shift in the peak of $v^5 f_1$ due to the velocity-dependent nature of e-e collisions. This is apparent in figure \ref{fig:ee_vs_ei}(b), with the peak of $v^5 f_1$ remaining at $v/v_{th}=2.63$ when the Epperlein \& Short approximation is utilized; the velocity of the heat-carrying electrons is unchanged. When the full isotropic e-e collision operator is used, this instead drops to $v/v_{th}=2.54$. This will become important for subsequent paper in this series studying non-locality\cite{walsh_paper1}, which finds that low-$Z$ plasmas are less susceptible to kinetic effects due to their heat-flux being carried by lower velocity (more collisional) electrons.
	
	Figure \ref{fig:ee_vs_ei}(c) also shows the other coefficients of this paper: their calculated value when the ES approximation is used, normalized to their value when using the full collision operator. While thermal conductivity is well reproduced across $Z$, the other coefficients are spuriously high at low $Z$, where e-e collisions are most important.
	
	\bibliographystyle{aipnum4-2}
	\bibliography{library}

\end{document}